\documentclass[superscriptaddress,amsmath, nofootinbib]{revtex4}
\usepackage{amsfonts}
\usepackage{graphicx}
\usepackage[latin1]{inputenc}
\usepackage{amsmath}
\usepackage{amssymb}
\usepackage{lineno}

\begin{document}


\title{An asymmetrical body: example of analytical solution for the rotation matrix in elementary functions and Dzhanibekov effect.}

\author{Alexei A. Deriglazov }
\email{alexei.deriglazov@ufjf.br} \affiliation{Depto. de Matem\'atica, ICE, Universidade Federal de Juiz de Fora,
MG, Brazil}

\date{\today}

\begin{abstract} 
We solved the Poisson equations, obtaining their exact solution in elementary functions for the rotation matrix of a free asymmetrical body with an angular velocity vector lying on separatrices. This allows us to discuss the temporal evolution of Dzhanibekov's nut directly in the laboratory system, where it is observed. The rotation matrix depends on two parameters with clear physical interpretation as a frequency and a damping factor of the solution. Qualitative analysis of the solution shows that it properly describes a single-jump Dzhanibekov effect. 
\end{abstract}

\maketitle 



\section{Introduction.}

Dzhanibekov's nut is a free asymmetrical body rotating around the axis very close to the intermediate axis of inertia of the body. Being in the state of an apparently stationary rotation, at regular time intervals such a body makes a sudden jump, abruptly changing the direction of rotation vector by 180 degrees. This paradoxical behavior was first observed by V. A. Dzhanibekov in 1985 during his trip to outer space and bears his name. A similar behavior of a body on the Earth's surface (that is in a gravitational field) is called the tennis racket effect. The clue for a theoretical explanation for this phenomenon is contained in Euler equations, which show that stationary rotation around the intermediate axis 
is non-stable \cite{{Whit_1917, Mac_1936, Lei_1965, Landau_8, Lev_2023, Ham_22}}.  Various aspects of the behavior of such a body, predicted from  Euler equations, are intensively discussed in the literature \cite{Chi_91, Kin_1992, Pet_2013, Zhyr_2020, Mur_21, Sai_20, Pet_21, Kor_23, Triv_20}.

However, solutions to Euler's equations describe the evolution of the instantaneous axis of rotation (that is the angular velocity) relative to a body-fixed frame, and thus contain only indirect information about the character of movement in the laboratory system, in which the body is observed. To understand the movement of a body in space, we need to find its rotation matrix, the latter is guided by Poisson equations. That is, in addition to the Euler equations, we also need to analyze the Poisson equations.

Poisson equations contain the angular velocity, which must first be found from Euler equations. Unfortunately, solutions to Euler equations generally can be written only in the form of elliptic integrals, which are not very illuminative. However, under special initial data, there are two (non-trivial) particular solutions in elementary functions \cite{Landau_8}, and it is they that turn out to be closely related to the Dzhanibekov effect. These two trajectories of angular velocity vectors are called separatrices. Then the natural question is whether it is also possible to solve the Poisson equations in elementary functions for this case? In this work we give an affirmative answer to this question, obtaining the explicit form of the rotation matrix of an asymmetrical body with angular velocity vector moving along the separatrices. 

Temporal evolution of the rotation axis along a separatrix is not a periodic function \cite{Landau_8}. As we will see, the same is true for the rotation matrix. In this case, an ideal body experiences a sudden jump only once: let it be launched with the rotation axis lying on a separatrix, and almost antiparallel to the conserved angular momentum. Then after some time, it will make a sudden jump so that the rotation axis becomes almost parallel to the angular momentum, after which it will asymptotically approach it\footnote{The body launched with the rotation axis almost parallel to the angular momentum, will approach it without any jump.}. This is the physical content of the rotation matrix that we found. The trajectories lying near separatrix are time-periodic functions \cite{Landau_8}. So it is expected that an ideal body along such a trajectory will experience multiple jumps at regular intervals, that is the Dzhanibekov effect. 

The work is organized as follows. In Sect. \ref{Dh_2} we describe the necessary notation and present the basic equations to be analyzed. 
In Sect. \ref{Dh_3} we solve the Poisson equations obtaining the rotation matrix in elementary functions, see Eq. (\ref{D25}). 
In the concluding section \ref{Dh_4} we present a qualitative analysis of the solution, showing that it properly describes a single-jump Dzhanibekov effect.
The rotation matrix will be constructed without assuming any kind of parametrization like Euler angles. In the Appendix, we discuss some inaccuracies that are often made when using the Euler angles.

\section{Notation and basic equations of the  problem.}\label{Dh_2}
Consider a free asymmetrical body with the principal inertia moments $I_3>I_2>I_1$. To describe its rotational movement, we place the origin of the laboratory frame in the center of mass. Besides, at the initial instant of time, say $t=0$, the laboratory axes (with unit vectors ${\bf e}_i$) should be chosen in the directions of the inertia axes (with unit vectors ${\bf R}_i(t)$), the latter we assume to be the body-fixed frame. Then the temporal evolution of the body can be obtained by solving the Euler-Poisson equations\footnote{We use the notation from \cite{AAD_2023_9, AAD_23}. In particular, by $({\bf a}, {\bf b})=a_i b_i$ and $[{\bf a}, {\bf b}]_i=\epsilon_{ijk}a_j b_k$ we denote the scalar and vector products of the vectors ${\bf a}$ and ${\bf b}$. $\epsilon_{ijk}$ is Levi-Civita symbol in three dimensions, with $\epsilon_{123}=1$.}
\begin{eqnarray}
I\dot{\boldsymbol\Omega}=[I{\boldsymbol\Omega}, {\boldsymbol\Omega}], \qquad ~ \label{D1} \\   
\dot R_{ij}=-\epsilon_{jkm}\Omega_k R_{im},  \label{D2}
\end{eqnarray}
with the inertia tensor of a diagonal form: $I_{ij}=diagonal(I_1, I_2, I_3)$. The basic variables in these $3+9$ equations are the rotation matrix $R_{ij}(t)$ and the components $\Omega_i(t)$ of angular-velocity vector ${\boldsymbol\omega}(t)$ relative to the body-fixed 
frame: ${\boldsymbol\omega}(t)\equiv\omega_i(t){\bf e}_i=\Omega_i(t){\bf R}_i(t)$. The initial data for the problem are $R_{ij}(0)=\delta_{ij}$, $\Omega_i(0)=\tilde\Omega_i$, $\tilde\Omega_i\in{\mathbb R}$. Any solution $R_{ij}(t)$ of the system (\ref{D1}), (\ref{D2}) with these initial data is an orthogonal matrix at any $t$. Besides, columns of the rotation matrix are just the basis vectors of the body-fixed frame: $R(t)=({\bf R}_1(t), {\bf R}_2(t), {\bf R}_3(t))$. 

For the latter use, we recall that velocity of rotation can be described using four different quantities. They are instantaneous angular velocity $\omega_i$, its components in body-fixed frame $\Omega_i$, conserved angular momentum $m_i$, and its components $M_i$ in a body-fixed frame. They are related as follows \cite{AAD_23}: 
\begin{eqnarray}\label{D3}
{\bf m}=R{\bf M}=RIR^T{\boldsymbol\omega}=RI{\bf\Omega}.   
\end{eqnarray}
At $t=0$ we get 
\begin{eqnarray}\label{D4}
{\bf m}={\bf M}(0)=I{\boldsymbol\omega}(0)=I{\bf\Omega}(0).   
\end{eqnarray}

\begin{figure}[t] \centering
\includegraphics[width=14cm]{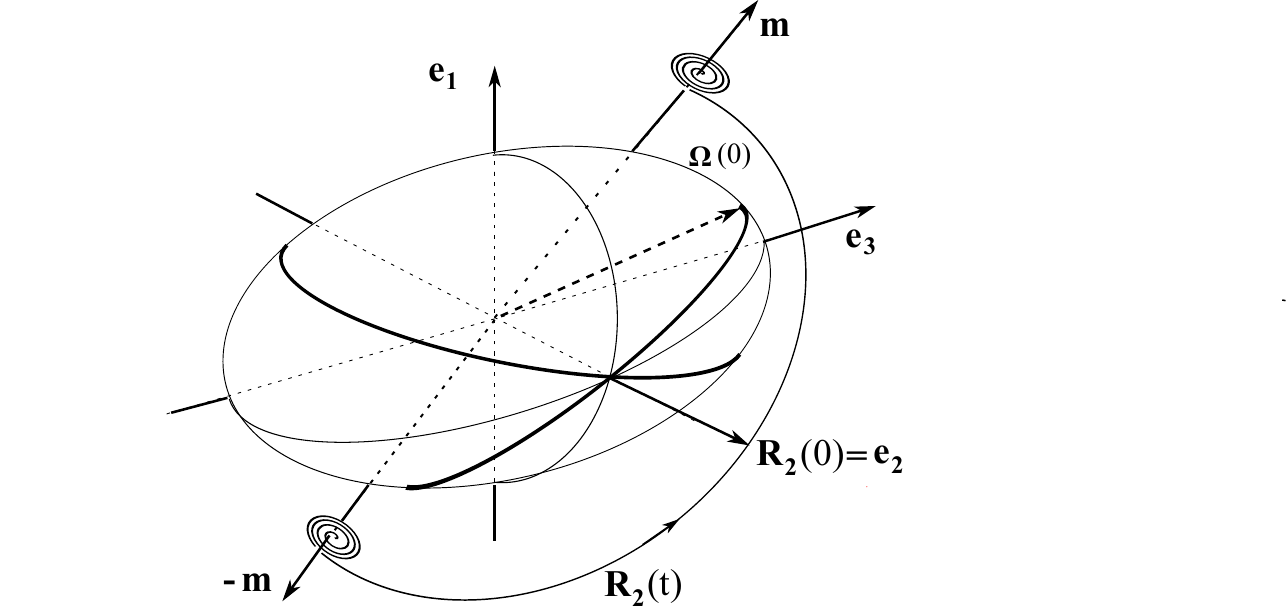}
\caption{E-ellipsoid is drawn at $t=0$. Two separatrices are drawn with bold lines. The curve denoted ${\bf R}_2(t)$ shows the trajectory of the intermediate axis for a single-jump Dzhanibekov effect.}\label{Dzh}
\end{figure}

{\bf Two particular solutions to Euler equations in elementary functions.} Euler equations (\ref{D1}) have two integrals of motion, which are the conserved energy and square of angular momentum
\begin{eqnarray}\label{D5}
I_1\Omega_1^2+I_2\Omega_2^2+I_3\Omega_3^2=2E, \qquad I_1^2\Omega_1^2+I_2^2\Omega_2^2+I_3^2\Omega_3^2={\bf m}^2.   
\end{eqnarray}
Hence for any solution ${\boldsymbol\Omega}(t)$ of Euler equations, the end of this vector at each $t$ lies on the curve that is an intersection of two ellipsoids (\ref{D5}). If we consider movements with fixed energy $E$, the intersections are possible when $2EI_1\le{\bf m}^2\le2EI_3$. When ${\bf m}^2$ is near $2EI_1$ or $2EI_3$, the intersections are closed curves around the vertices of the largest ${\bf R}_3(t)$ and smallest ${\bf R}_1(t)$ axis. These regions are separated by two curves with 
\begin{eqnarray}\label{D6}
{\bf m}^2=2EI_2. 
\end{eqnarray}
They are called separatrices. They pass through vertices of the intermediate axis ${\bf R}_2(t)$ of the E-ellipsoid, see Figure \ref{Dzh}. 
Geometrically, they turn out to be two ellipses \cite{Landau_8}. 

All trajectories (\ref{D5}) intersect the plane $\Omega_2=0$ twice, so without loss of generality we can choose the initial data for the problem (\ref{D1})  on a half-plane as follows: ${\boldsymbol\Omega}(0)=(\tilde\Omega_1, 0, \tilde\Omega_3>0)^T$. Then Eqs. (\ref{D5}), 
taken at  at $t=0$, give the initial data through $E$ and ${\bf m}^2$ as follows: 
\begin{eqnarray}\label{D7}
\tilde\Omega_1^2=\frac{-{\bf m}^2+2EI_3}{I_1 I_2 I_{3-1}}, \qquad \tilde\Omega_3^2=\frac{{\bf m}^2-2EI_1}{I_3 I_2 I_{3-1}}, \quad 
\mbox{where} \quad I_{3-1}\equiv\frac{I_3-I_1}{I_2}, \quad I_{2-1}\equiv\frac{I_2-I_1}{I_3},\quad \mbox{and so on}. 
\end{eqnarray}
Using Eqs. (\ref{D5}) at $t=0$ together with (\ref{D6}) in the expressions (\ref{D7}), we get the initial data for selecting the separatrices, which are thereby determined by $\tilde\Omega_3$ as follows:  
\begin{eqnarray}\label{D8}
{\boldsymbol\Omega}(0)=(\epsilon a\tilde\Omega_3, ~ 0, ~ \tilde\Omega_3>0)^T, \qquad \epsilon=\pm 1, \qquad 
a\equiv \sqrt{\frac{I_{3-2}}{I_{2-1}}}. 
\end{eqnarray}

Resolving the algebraic system (\ref{D5}) we get
\begin{eqnarray}\label{D5.1}
\Omega_1^2=-\frac{{\bf m}^2-2EI_2}{I_1 I_3 I_{2-1}}+\frac{I_{3-2}}{I_{2-1}}\Omega_3^2, \qquad 
\Omega_2^2=\frac{{\bf m}^2-2EI_1}{I_2 I_3 I_{2-1}}-\frac{I_{3-1}}{I_{2-1}}\Omega_3^2. \nonumber
\end{eqnarray}
Taking into account the equations (\ref{D6}) and (\ref{D7}) in these expressions, we represent $\Omega_1(t)$ and $\Omega_2(t)$ through $\Omega_3(t)$ as follows: 
\begin{eqnarray}\label{D9}
\Omega_1=\epsilon a\Omega_3, \qquad \Omega_2=\pm b\sqrt{\tilde\Omega_3^2-\Omega_3^2}, \qquad \mbox{where} \quad  b\equiv\sqrt{\frac{I_{3-1}}{I_{2-1}}}. 
\end{eqnarray}
To be specific, we will continue our analysis for the separatrix with $\epsilon=+1$.
Let us consider the third component of the vector equation (\ref{D1}): $I_3\dot\Omega_3=(I_1-I_2)\Omega_1\Omega_2$, or, equivalently, 
$\dot\Omega_3=I_{1-2}\Omega_1\Omega_2$. Substituting (\ref{D9}) into the right side of this equation, we obtain the following closed first-order equation for determining $\Omega_3(t)$: $\dot\Omega_3=\pm\sqrt{I_{3-2}I_{3-1}}~\Omega_3\sqrt{\tilde\Omega_3^2-\Omega_3^2}$. Let us discuss the choice of the sign in this equation.  When $t=0$, its solution $\Omega_3(t)$ should reach its maximum value $\Omega_3(0)=\tilde\Omega_3$, see Figure \ref{Dzh}. This implies that $\dot\Omega_3(t)>0$ in the region $t<0$, while $\dot\Omega_3(t)<0$ in the region $t>0$. Therefore, as the equation of the separatrix in the region $t<0$ we take  
\begin{eqnarray}\label{D10}
\dot\Omega_3=+\sqrt{I_{3-2}I_{3-1}}~\Omega_3\sqrt{\tilde\Omega_3^2-\Omega_3^2}.
\end{eqnarray}
In the region $t>0$ the separatrix is described by this equation with the minus sign on the right side. 
Separating the variables in Eq. (\ref{D10}): $\frac{d\Omega_3}{\Omega_3\sqrt{\tilde\Omega_3^2-\Omega_3^2}}=\sqrt{I_{3-2}I_{3-1}}dt$, it is immediately integrated as follows: $\frac{-1}{2\tilde\Omega_3}\left(ln\left[\tilde\Omega_3+\sqrt{\tilde\Omega_3^2-\Omega_3^2}\right]-ln\left[\tilde\Omega_3-\sqrt{\tilde\Omega_3^2-\Omega_3^2}\right]\right)=\sqrt{I_{3-2}I_{3-1}}t+c$. Using the initial data $\Omega_3(0)=\tilde\Omega_3$ we conclude $c=0$, then 
\begin{eqnarray}\label{D11}
\frac{\tilde\Omega_3+\sqrt{\tilde\Omega_3^2-\Omega_3^2}}{\tilde\Omega_3-\sqrt{\tilde\Omega_3^2-\Omega_3^2}}=e^{-2\omega t}, \qquad \omega\equiv\sqrt{I_{3-2}I_{3-1}}~\tilde\Omega_3.
\end{eqnarray}
Separating $\Omega_3$ from this equality  we get
\begin{eqnarray}\label{D12}
\Omega_3(t)=\frac{\tilde\Omega_3}{\cosh\omega t}. 
\end{eqnarray}
This implies $\sqrt{\tilde\Omega_3^2-\Omega_3^2(t)}=-\tilde\Omega_3\tanh\omega t$ in the region $t<0$. Taking this into account, the function (\ref{D12}) obeys the equation (\ref{D11}).  Repeating these calculations in the region $t>0$, we arrive at the same final expression (\ref{D12}), which thereby represents the entire separatrix.  

Putting the solution (\ref{D12}) into Eq. (\ref{D9}), we obtain the remaining components up to a sign. The resulting vectors ${\boldsymbol\Omega}(t)$, for which $\Omega_1(t)$ and $\Omega_2(t)$ have the same sign, turn out to be solutions to Euler equations (\ref{D1}).   So the final expression for the time evolution of angular velocity along the separatrices is
\begin{eqnarray}\label{D13}
{\boldsymbol\Omega}(t)=\left(\frac{\epsilon a\tilde\Omega_3}{\cosh\omega t}, \quad \epsilon b\tilde\Omega_3 \tanh\omega t, \quad \frac{\tilde\Omega_3}{\cosh\omega t}\right)^T, \quad \epsilon=\pm 1, \quad a\equiv \sqrt{\frac{I_{3-2}}{I_{2-1}}}, \quad 
b\equiv\sqrt{\frac{I_{3-1}}{I_{2-1}}}, \quad \omega\equiv\sqrt{I_{3-2}I_{3-1}}~ \tilde\Omega_3. 
\end{eqnarray}
The evolution is not a periodic function of time: ${\boldsymbol\Omega}(\pm\infty)=(0, ~ \pm\epsilon b\tilde\Omega_3, ~ 0 )^T$. This equality means that the vector ${\boldsymbol\Omega}(t)$  in the body-fixed frame ${\bf R}_i$ asymptomatically approaches to the intermediate axis ${\bf R}_2$. 

The following identities (for $\epsilon=+1$):
\begin{eqnarray}\label{D13.1}
\Omega_1=a\Omega_3, \qquad \frac{d}{dt}\left(\frac{1}{\Omega_3}\right)=-\frac{I_{1-2}a\Omega_2}{\Omega_3}, 
\qquad b\omega=I_{3-1}a\tilde\Omega_3,
\end{eqnarray}
will be repeatedly used in subsequent calculations.

\section{Solution to Poisson equations in elementary functions.}\label{Dh_3} 

Denoting the rows of rotation matrix $R_{ij}$ as ${\bf G}_1, {\bf G}_2$ and ${\bf G}_3$, the original system of 9 Poisson equations (\ref{D2}) splits into three systems: $\dot{\bf G}_i=[{\bf G}_i, {\boldsymbol\Omega}]$.  So, denoting any one row ${\bf G}_i$ by ${\boldsymbol\gamma}$, we need to solve the following equations: $\dot{\boldsymbol\gamma}=[{\boldsymbol\gamma}, {\boldsymbol\Omega}(t)]$, or, in components  
\begin{eqnarray}\label{D14} 
\dot\gamma _1=\gamma_2\Omega_3-\gamma_3\Omega_2, \qquad \dot\gamma _2=\gamma_3\Omega_1-\gamma_1\Omega_3, 
\qquad \dot\gamma _3=\gamma_1\Omega_2-\gamma_2\Omega_1, 
\end{eqnarray}
with known functions $\Omega_i(t)$ written in (\ref{D13}). To be specific, we work with only one separatrix,  with $\epsilon=+1$. These equations have two integrals of motion. The first is ${\boldsymbol\gamma}^2=\mbox{const}$. The second follows from conservation of angular momentum:  $m_k=(RI{\boldsymbol\Omega})_k=\sum_{i}I_i\Omega_i(G_k)_i$, and reads as follows:
\begin{eqnarray}\label{D15} 
I_1\Omega_1\gamma_1+I_2\Omega_2\gamma_2+I_3\Omega_3\gamma_3=c. 
\end{eqnarray}
The whole rotation matrix will be restored, choosing three particular solutions to the system (\ref{D14}) and (\ref{D15}). ${\bf G}_1(t)$ is the solution ${\boldsymbol\gamma}(t)$ that obeys the initial data ${\boldsymbol\gamma}(0)=(1, 0, 0)$, and taking  $c=m_1=(I{\boldsymbol\Omega}(0))_1=I_1a\tilde\Omega_3$ according to (\ref{D15}), (\ref{D4}) and (\ref{D13}).  ${\bf G}_2(t)$ is the solution ${\boldsymbol\gamma}(t)$ that obeys the initial data ${\boldsymbol\gamma}(0)=(0, 1, 0)$ and taking  $c=m_2=0$. At last,  ${\bf G}_3(t)$ is the solution ${\boldsymbol\gamma}(t)$ that obeys the initial data ${\boldsymbol\gamma}(0)=(0, 0, 1)$ and taking $c=m_3=I_3\tilde\Omega_3$. 

Using (\ref{D13.1}) in (\ref{D14}) and (\ref{D15}) we get 
\begin{eqnarray}\label{D16} 
\dot\gamma _1=\gamma_2\Omega_3-\gamma_3\Omega_2, ~ ~  \label{D16.1} \\
\dot\gamma _2=a\gamma_3\Omega_3-\gamma_1\Omega_3, \label{D16.2} \\
\qquad \dot\gamma _3=\gamma_1\Omega_2-a\gamma_2\Omega_3, \label{D16.3}  \\
I_1a\gamma_1+I_3\gamma_3=\frac{1}{\Omega_3}(c-I_2\Omega_2\gamma_2).  \label{D16.4}
\end{eqnarray}
Solving the equations (\ref{D16.2}) and (\ref{D16.4}) with respect to $\gamma_1$ and $\gamma_3$ (with the use of the identity $I_1 a^2+I_3=I_2 b^2$)
\begin{eqnarray}\label{D17}
\gamma_1=\frac{1}{I_2 b^2\Omega_3}\left[ac-aI_2\Omega_2\gamma_2-I_3\dot\gamma_2\right], \qquad 
\gamma_3=\frac{1}{I_2 b^2\Omega_3}\left[c-I_2\Omega_2\gamma_2+I_1a\dot\gamma_2\right],
\end{eqnarray}
we substitute them into (\ref{D16.1}). By direct calculations with the use of Eqs. (\ref{D13}) and (\ref{D13.1}), we get the following closed equation for determining $\gamma_2(t)$: 
\begin{eqnarray}\label{D18}
\frac{d^2}{dt^2}\left(\frac{\gamma_2}{\Omega_3}\right)+\left[\Omega_2^2+\frac{I_{3-1}}{I_{2-1}}\Omega_3^2\right]\frac{\gamma_2}{\Omega_3}=\frac{I_{3-2}+1}{I_3}\frac{c\Omega_2}{\Omega_3}.
\end{eqnarray}
Remarkably, substituting $\Omega_2$ and $\Omega_3$ from (\ref{D13}), we obtain the equation of a harmonic oscillator for the 
quantity $\frac{\gamma_2}{\Omega_3}$, with the time-independent proper 
frequency $k$, and under the influence of an external time-dependent force
\begin{eqnarray}\label{D19}
\frac{d^2}{dt^2}\left(\frac{\gamma_2}{\Omega_3}\right)+k^2\frac{\gamma_2}{\Omega_3}=
\frac{I_3-I_2+I_1}{I_1I_3}cb\sinh\omega t, \qquad k\equiv b\tilde\Omega_3. 
\end{eqnarray}
Its general solution is
\begin{eqnarray}\label{D20}
\gamma_2(t)=\Omega_3(t)\left[A\cos(kt+k_0)+\frac{cb}{I_2 k^2}\sinh\omega t\right]. 
\end{eqnarray}
Using this in Eq. (\ref{D17}), and presenting the resulting expressions in terms of initial value $\tilde\Omega_3$, the general solution to equations 
(\ref{D16.1})-(\ref{D16.4}) with integration constants $A$ and $k_0$ reads as follows:
\begin{eqnarray}\label{D21}
\gamma_1(t)=\frac{1}{I_2b^2}\left[b\tilde\Omega_3I_3A\sin(kt+k_0)-b\tilde\Omega_3 aI_1A\cos(kt+k_0)\tanh\omega t+
\frac{I_1 ac}{I_2\tilde\Omega_3}\frac{1}{\cosh\omega t}\right], \cr
\gamma_2(t)=\frac{\tilde\Omega_3}{\cosh\omega t}A\cos(kt+k_0)+\frac{c}{I_2 b\tilde\Omega_3}\tanh\omega t, \qquad \qquad \qquad \qquad \qquad \qquad \qquad  \qquad \cr
\gamma_3(t)=\frac{1}{I_2b^2}\left[-b\tilde\Omega_3 aI_1A\sin(kt+k_0)-b\tilde\Omega_3 I_3A\cos(kt+k_0)\tanh\omega t+
\frac{I_3 c}{I_2\tilde\Omega_3}\frac{1}{\cosh\omega t}\right]. 
\end{eqnarray}
At the initial instant $t=0$ we get
\begin{eqnarray}\label{D22}
\gamma_1(0)=\frac{1}{I_2b^2}\left[b\tilde\Omega_3I_3A\sin k_0+\frac{I_1 ac}{I_2\tilde\Omega_3}\right], \qquad 
\gamma_2(0)=\tilde\Omega_3 A\cos k_0, \qquad 
\gamma_3(0)=\frac{1}{I_2b^2}\left[-b\tilde\Omega_3 aI_1A\sin k_0+\frac{I_3 c}{I_2\tilde\Omega_3}\right]. 
\end{eqnarray}
As we saw above, the three rows of the matrix $R_{ij}$ are obtained from Eqs. (\ref{D21}) if we adjust the integration constants $A$ and $k_0$ with initial data as indicated below Eq. (\ref{D15}). This can equally be done with the use of Eqs. (\ref{D22}) instead of (\ref{D21}). Solving the equations (\ref{D22}) with these data, we get, in each case 
\begin{eqnarray}\label{D23}
A=\frac{I_3}{I_2 b\tilde\Omega_3}, \quad k_0=\frac{\pi}{2}; \qquad \quad
A=\frac{1}{\tilde\Omega_3}, \quad k_0=0; \qquad \quad
A=-\frac{I_1 a}{I_2 b\tilde\Omega_3}, \quad k_0=\frac{\pi}{2}. 
\end{eqnarray}
Substituting these values into Eqs. (\ref{D21}),  we get the rotation matrix $R$. The most compact expression for it appears if instead of the parameters $a$ and $b$ we introduce unit vector $\hat{\bf m}$ in the direction of conserved angular momentum ${\bf m}$. According to Eqs. (\ref{D4}) and (\ref{D8}), it is expressed in terms of $a$ and $b$ as follows: $\hat{\bf m}=(\hat m_1, 0, \hat m_3)^T\equiv\frac{{\bf m}}{|{\bf m}|}=\frac{I{\boldsymbol\Omega}(0)}{|I{\boldsymbol\Omega}(0)|}=(\frac{I_1 a}{I_2 b}, ~ 0, ~ \frac{I_3}{I_2 b})^T$. In terms of the moments of  inertia it is
\begin{eqnarray}\label{D24}
\hat{\bf m}=\left(\frac{I_1}{I_2}\sqrt{\frac{I_{3-2}}{I_{3-1}}}, ~ 0, ~ \frac{I_3}{I_2}\sqrt{\frac{I_{2-1}}{I_{3-1}}}~\right)^T. 
\end{eqnarray}
With these notation, the rotation matrix $R^{+}$ for the movement along the separatrix with $\epsilon=+1$ reads as follows
\begin{eqnarray}\label{D25}
\left(
\begin{array}{ccc}
\frac{\hat m_1^2}{\cosh\omega t}+\hat m_3^2\cos kt+\hat m_1\hat m_3\sin kt\tanh\omega t & 
\frac{-\hat m_3\sin kt}{\cosh\omega t}+\hat m_1\tanh\omega t & 
\hat m_1\hat m_3(\frac{1}{\cosh\omega t}-\cos kt)+\hat m_3^2\sin kt\tanh\omega t \vspace{4mm} \\ 
\hat m_3\sin kt-\hat m_1\cos kt\tanh\omega t & \frac{\cos kt}{\cosh\omega t} & -\hat m_1\sin kt-\hat m_3\cos kt\tanh\omega t 
\vspace{4mm}  \\
\hat m_1\hat m_3(\frac{1}{\cosh\omega t}-\cos kt)-\hat m_1^2\sin kt\tanh\omega t &
\frac{\hat m_1\sin kt}{\cosh\omega t}+\hat m_3\tanh\omega t & 
\frac{\hat m_3^2}{\cosh\omega t}+\hat m_1^2\cos kt-\hat m_1\hat m_3\sin kt\tanh\omega t
\end{array}\right) ~ 
\end{eqnarray}

Rotation matrix $R^{-}$ for the separatrix with $\epsilon=-1$ is obtained from this expression replacing $\hat m_1$ on $-\hat m_1$.
As it should be, the obtained matrix is orthogonal at any $t$, in particular $R_{ij}(0)=\delta_{ij}$. 
The final expression admits the limit of a symmetrical body with $I_3=I_2>I_1$. This could be a fairly prolated cylinder of uniform density.
In this limit Eq. (\ref{D25}) turn into the expected expression 
\begin{eqnarray}\label{D26}
\left(
\begin{array}{ccc}
\cos\tilde\Omega_3 t & -\sin\tilde\Omega_3 t & 0 \\
\sin\tilde\Omega_3 t & \cos\tilde\Omega_3 t & 0 \\
0 & 0 & 1
\end{array}\right),  
\end{eqnarray}
that describe a rotation of the cylinder around its transverse axis $\hat{\bf m}=(0, 0, 1)$. 

\begin{figure}[t] \centering
\includegraphics[width=10cm]{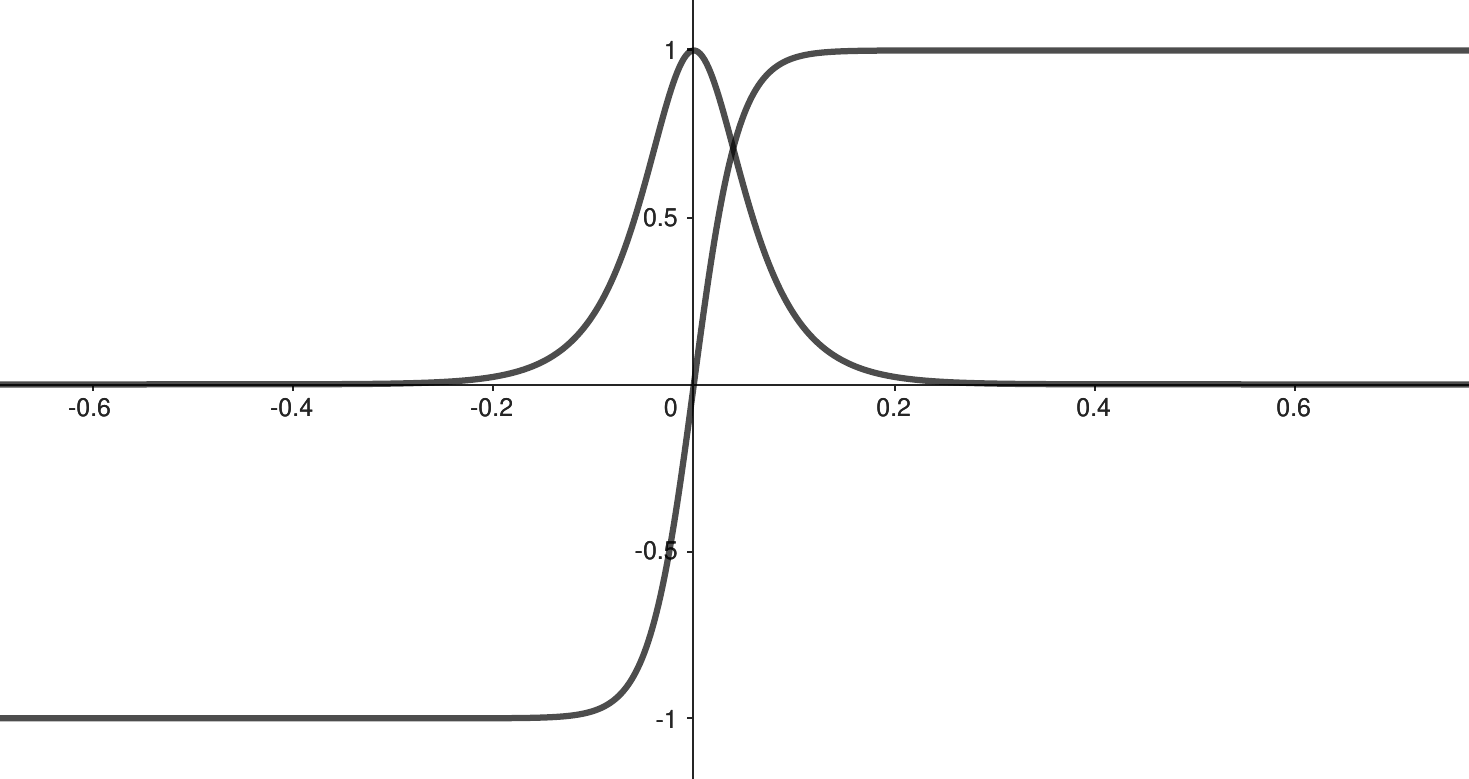}
\caption{Graphs of the damping functions $\cosh^{-1}(22.423 t)$ and $\tanh(22.423 t)$.}\label{cosh}
\end{figure}

The movement (\ref{D25}) is determined by the frequency
\begin{eqnarray}\label{D27}
k\equiv \sqrt{\frac{I_{3-1}}{I_{2-1}}}~ \tilde\Omega_3,
\end{eqnarray}
and by the damping factor
\begin{eqnarray}\label{D28}
\omega\equiv\sqrt{I_{3-2}I_{3-1}}~ \tilde\Omega_3. 
\end{eqnarray}
To estimate the latter, let us consider the homogeneous rectangular plate of sides $a=7$ cm., $b=4$ cm., $c=2$ cm.,  and 
with $\tilde\Omega_3=2\pi\times 5$ rad/sec. Its inertia moments are \cite{Landau_8}: $I_3=\frac{\mu}{12}(a^2+b^2)=\frac{\mu}{12}65$, $I_2=\frac{\mu}{12}(a^2+c^2)=\frac{\mu}{12}53$ and $I_1=\frac{\mu}{12}(b^2+c^2)=\frac{\mu}{12}20$. Then $\cosh^{-1}(\omega t)=\cosh^{-1}(22.423 t)$ and $\tanh(\omega t)=\tanh(22.423 t)$.  In particular, at $t=0.3$ sec. we get  $\cosh^{-1}(22.423\times 0.3)=0,0023$ 
and $\tanh(22.423\times 0.3)=0.999997$. The graphs of these functions are presented in Figure \ref{cosh}. 
The function $\cosh^{-1}(\omega t)$ makes a sharp jump on the interval $t\in[-0.3, ~ 0.3]$, and is practically equal to zero outside it. The 
function $\tanh(\omega t)$ also makes a sharp jump approximately on the same interval and is practically equal to $\pm 1$ outside it. 

In the limit $t\rightarrow\pm\infty$ the rotation matrix (\ref{D25}) acquires the form
\begin{eqnarray}\label{D29}
R^{+}_{ij}(\pm\infty)=\left(
\begin{array}{ccc}
\hat m_3^2\cos kt \pm\hat m_1\hat m_3\sin kt ~ & \pm\hat m_1 & ~ -\hat m_1\hat m_3\cos kt \pm\hat m_3^2\sin kt \vspace{4mm} \\ 
\hat m_3\sin kt\mp\hat m_1\cos kt ~  & 0 & ~ -\hat m_1\sin kt\mp\hat m_3\cos kt  \vspace{4mm}  \\
-\hat m_1\hat m_3\cos kt \mp\hat m_1^2\sin kt ~ & \pm\hat m_3 & ~ \hat m_1^2\cos kt \mp\hat m_1\hat m_3\sin kt
\end{array}\right).  
\end{eqnarray}
This is the rotation of the frequency $k$ around the vector of conserved angular momentum $\pm{\bf m}$. That is in this limit the intermediate 
axis ${\bf R}_2(t)$ occupies the position of $\pm{\bf m}$, while other two axes rotate around it.

\section{Conclusion. Qualitative discussion of the rotation matrix and Dzhanibekov effect.}\label{Dh_4}

We solved Poisson equations (\ref{D2}) for an asymmetrical body with an angular velocity vector lying on a separatrix, obtaining the expression (\ref{D25}) for the rotation matrix in elementary functions.  The resulting matrix depends on two parameters which determine the frequency and asymptotic damping of the solution at $t\rightarrow\pm\infty$. The motion guided by this matrix has a simple physical meaning: the body in the process of motion tends to superpose its axis of instantaneous rotation with the direction of a conserved vector of angular momentum. To see 
this we consider, for definiteness, the homogeneous rectangular plate described below Eq. (\ref{D28}). Consider the plate that at $t=0$ has angular 
velocity ${\boldsymbol\Omega}(0)=(\epsilon a\tilde\Omega_3, ~ 0, ~ \tilde\Omega_3>0)^T$, $\tilde\Omega_3=2\pi\times 5$ rad/sec. According to (\ref{D4}), the conserved angular momentum of the plate is ${\bf m}=(I_1 a\tilde\Omega_3, ~ 0, ~I_3\tilde\Omega_3)$. Both vectors lie in the plane of laboratory vectors ${\bf e}_1$, ${\bf e}_3$, see Figure \ref{Dzh}. 
The intermediate axis ${\bf R}_2(t)$ of the plate is the second column of the rotation matrix (\ref{D25}). Let us analyze its time evolution during the interval $t\in [-t_1, t_1]$ where $t_1>>1$. During the interval $t\in[-t_1,    -0.3]$ we can approximate the axis as follows:
\begin{eqnarray}\label{D30}
{\bf R}_2(t)=\left(\hat m_1\tanh\omega t-\frac{\hat m_3\sin kt}{\cosh\omega t}, ~~ \frac{\cos kt}{\cosh\omega t}, ~ ~ \hat m_3\tanh\omega t+\frac{\hat m_1\sin kt}{\cosh\omega t}\right)^T\approx \cr 
\left(-\hat m_1-0.002\hat m_3\sin kt, ~ ~  0.002\cos kt, ~ ~ -\hat m_3+0.002\hat m_1\sin kt\right)^T. \qquad 
\end{eqnarray}
This means that the intermediate axis moves in the vicinity of the vector $-\hat{\bf m}=(-\hat m_1, 0, -\hat m_3)$ in an expanding spiral with a very small step, see Figure \ref{Dzh}. Next, during the interval $t\in[-0.3, ~ 0.3]$ of $0.6$ sec., the intermediate axis abruptly changes its position (see Figure \ref{cosh}), ending up in the vicinity of the vector $\hat{\bf m}=(\hat m_1, 0, \hat m_3)$. During the interval $t\in[0.3,~  t_1]$ we can approximate the axis as follows: 
\begin{eqnarray}\label{D31}
{\bf R}_2(t)=\left(\hat m_1\tanh\omega t-\frac{\hat m_3\sin kt}{\cosh\omega t}, ~~ \frac{\cos kt}{\cosh\omega t}, ~ ~ \hat m_3\tanh\omega t+\frac{\hat m_1\sin kt}{\cosh\omega t}\right)^T\approx \cr  
\left(\hat m_1-0.002\hat m_3\sin kt, ~ ~  0.002\cos kt, ~ ~ \hat m_3+0.002\hat m_1\sin kt\right)^T. \qquad 
\end{eqnarray}
This means that the intermediate axis asymptotically approaches the conserved angular momentum $\hat{\bf m}=(\hat m_1, 0, \hat m_3)$ in a tapering spiral. 

In resume, our solution (\ref{D25}) for the rotation matrix in elementary functions, which corresponds to the non-periodic movement of angular velocity along a separatrix, describes a single-jump Dzhanibekov effect. 

The trajectories lying near the separatrix are described with time-periodic functions \cite{Landau_8}. In this case, we could expect that on the right side of the equation (\ref{D19}) of a harmonic oscillator will appear a periodic external force instead of the damping force $\sinh\omega t$. This implies a periodic rotation matrix as a solution. So, we expect a periodic solution with properties similar to those discussed above: an ideal body along such a trajectory will experience multiple jumps at regular intervals, that is the Dzhanibekov effect.

\begin{acknowledgments}
The work has been supported by the Brazilian foundation CNPq (Conselho Nacional de
Desenvolvimento Cient\'ifico e Tecnol\'ogico - Brasil). 
\end{acknowledgments}

\section{Appendix}\label{Dh_5}

The rotation matrix has been constructed above without assuming any kind of parametrization like Euler angles. The rotation matrix in terms of Euler angles reads as follows: 
\begin{eqnarray}\label{D32}
R=\left(
\begin{array}{ccc}
\cos\psi\cos\varphi-\sin\psi\cos\theta\sin\varphi  &  -\sin\psi\cos\varphi-\cos\psi\cos\theta\sin\varphi &\sin\theta\sin\varphi \\
\cos\psi\sin\varphi+\sin\psi\cos\theta\cos\varphi & -\sin\psi\sin\varphi+\cos\psi\cos\theta\cos\varphi & -\sin\theta\cos\varphi \\
\sin\psi\sin\theta & \cos\psi\sin\theta & \cos\theta
\end{array}\right),
\end{eqnarray}
and then the problem reduces to the search for their dependence on time. To avoid possible confusion, here we discuss some peculiarities, that should be taken into account when working with the Euler angles.

Let us assume that we solved the Euler equations for some given initial data 
\begin{eqnarray}\label{D33}
{\boldsymbol\omega}(0)={\boldsymbol\Omega}(0)=(\tilde\Omega_1, \tilde\Omega_2, \tilde\Omega_3).
\end{eqnarray}
According to Eq.  (\ref{D3}), we can write the following vector 
equality with known left hand side: ${\boldsymbol\Omega}(t)=I^{-1}R^T(\theta(t), \varphi(t), \psi(t)){\bf m}$. In components we get 
\begin{eqnarray}\label{D34}
\Omega_1(t)=\frac{m_1}{I_1}[\cos\psi\cos\varphi-\sin\psi\cos\theta\sin\varphi]+\frac{m_2}{I_1}[\cos\psi\sin\varphi+\sin\psi\cos\theta\cos\varphi]+\frac{m_3}{I_1}\sin\psi\sin\theta, \cr
\Omega_2(t)=\frac{m_1}{I_2}[-\sin\psi\cos\varphi-\cos\psi\cos\theta\sin\varphi]+\frac{m_2}{I_2}[-\sin\psi\sin\varphi+\cos\psi\cos\theta\cos\varphi]+\frac{m_3}{I_2}\cos\psi\sin\theta,  \cr 
\Omega_3(t)=\frac{m_1}{I_3}\sin\theta\sin\varphi-\frac{m_2}{I_3}\sin\theta\cos\varphi+\frac{m_3}{I_3}\cos\theta. 
\end{eqnarray}
So it remains to resolve these algebraical equalities relative to the Euler angles. Let us try to do this in accordance with the following widely used reasoning. Before solving the equations (\ref{D34}), we simplify them as follows. We take the laboratory axis ${\bf e}_3$ in the direction of constant in space vector of angular momentum ${\bf m}$. Then ${\bf m}=(0, 0, M)$, where $M=|{\bf m}|$, and Eqs. (\ref{D34}) acquire the form 
\begin{eqnarray}\label{D35}
M\sin\psi\sin\theta=I_1\Omega_1(t)\equiv M_1, \qquad M\cos\psi\sin\theta=I_2\Omega_2(t)\equiv M_2, \qquad M\cos\theta=I_3\Omega_3(t)\equiv M_3,
\end{eqnarray}
where $M_i(t)$ are components of angular momentum in the body-fixed frame (these equalities should be compared with Eqs. (37.13) in \cite{Landau_8}). 
Using the previously obtained solutions $\Omega_i(t)$ to Euler equations, we immediately get two of the desired 
quantities: $\cos\theta(t)=I_3\Omega_3(t)/M$, $\tan\psi(t)=I_1\Omega_1(t)/I_2\Omega_2(t)$. 

To understand, what is wrong here, note the following. Since we are working with equations in which the inertia tensor is diagonal, then by choosing the third axis of laboratory in the direction of ${\bf m}$, we also have the third axis of inertia ${\bf R}_3(0)$ in the same direction. Besides, from $R_{ij}(0)=\delta_{ij}$ together with Eq. (\ref{D4}) we get 
\begin{eqnarray}\label{D36}
{\boldsymbol\omega}(0)=I^{-1}{\bf m}=(0, 0, M/I_3)\parallel{\bf R}_3(0). 
\end{eqnarray}
That is, under the assumptions made above, the initial angular velocity of the body turns out to be directed along the third axis of inertia. But in this case, the subsequent evolution of an ideal body is just a stationary rotation around this axis. It is clear, that calculations made using the 
equations (\ref{D35})  have nothing to do with this movement. Comparing (\ref{D33}) with (\ref{D36}), we conclude that this occurred because the Euler equations were solved in a coordinate system different from that chosen to simplify the Poisson equations (\ref{D34}). In other words,  
in writing the equations (\ref{D35}), the components of two vectors taken in two different coordinate systems were equalized. More detailed discussion around these points can be found in \cite{AAD_23} and \cite{AAD23_3}.


\begin{thebibliography}{99} 

\bibitem{Whit_1917} E. T. Whittaker, {\it A treatise on the analytical dynamics of particles and rigid bodies}, (Cambridge: at the University press, 1917). 

\bibitem{Mac_1936} W. D. MacMillan, {\it Dynamics of rigid bodies}, (Dover Publications Inc., New-York, 1936). 

\bibitem{Lei_1965} E. Leimanis, {\it The general problem of the motion of coupled rigid bodies about a fixed point}, (Springer-Verlag, 1965). 

\bibitem{Landau_8} L. D. Landau and E. M. Lifshitz, {\it Mechanics}, Volume 1, third edition, (Elsevier, 1976).

\bibitem{Lev_2023} F. L. Chernousko, L. D. Akulenko, D. D. Leshchenko, {\it Evolution of motions of a rigid body about its center of mass}, (Springer, 2017). 

\bibitem{Ham_22} H. M. Yehia, {\it Rigid body dynamics. A Lagrangian approach},  Advances in Mechanics and Mathematics, V. 45, Birkh\"auser,  2022. 

\bibitem{Chi_91} M. S. Ashbaugh, C. C. Chicone, R. H. Cushman, {\it The twisting tennis racket}, Dyn. Diff. Equat. {\bf 3} (1991) 67-85.

\bibitem{Kin_1992}  H. Kinoshita, {\it Analytical expansions of torque-free motions for short and long axis modes},  Celestial Mech. Dyn. Astron. {\bf 53} (1992) 365-375. 

\bibitem{Pet_2013} A. G. Petrov, S. E. Volodin, {\it Janibekov's effect and the laws of mechanics}, Dokl. Phys., {\bf 58}, N 8 (2013) 349-353.  

\bibitem{Zhyr_2020} V. F. Zhuravlev, G. M. Rozenblat, {\it Estimates of solutions during motion of the
Euler-Poinsot top and explanation of the experiment with Dzhanibekov's nut}, Rus. J. Nonlin. Dyn.,  {\bf 16} N 3, (2020) 517-525. 

\bibitem{Mur_21}  C. Murakami, {\it Analytical solution of the Euler-Poinsot problem}, J. Geom. Symmetry Phys. 60 (2021) 25-46; arXiv:2103.16056. 

\bibitem{Sai_20} S. Sailer, S. R. Eugster and R. I. Leine, {\it The tippedisk: a tippetop without rotational symmetry},  Regular and Chaotic Dynamics, {\bf  25} N 6 (2020) 553.  

\bibitem{Pet_21} C. Peterson, and W. Schwalm, {\it Euler's rigid rotators, Jacobi elliptic functions, and the Dzhanibekov or tennis 
racket effect}, American Journal of Physics {\bf 89} (2021) 349.  

\bibitem{Kor_23} V. V. Koryanov and A. S. Kukharenko, {\it  Influence of the inertia resultant moments inequality on the aerial vehicle rotational motion},  Engineering Journal: Science and Innovation, {\bf  2} (2023) 134. 

\bibitem{Triv_20} P. Trivailo and H. Kojima, {\it  Enhancement of the attitude dynamics capabilities of the spinning spacecraft using 
inertial morphing}, Aeronautica Journal, {\bf 124} 1276, (2020) 838-871.

\bibitem{AAD_2023_9} A. A. Deriglazov, {\it  Euler-Poisson equations of a dancing spinning top, integrability and
examples of analytical solutions},  Commun. Nonlinear Sci. Numer. Simulat. {\bf 127} (2023) 107579; arXiv:2307.12201. 

\bibitem{AAD_23} A. A. Deriglazov, {\it Lagrangian and Hamiltonian formulations of asymmetric rigid body, considered as a constrained system},  
Eur. J. Phys. {\bf 44} (2023) 065001;  arXiv:2301.10741.

\bibitem{AAD23_3} A. A. Deriglazov, {\it Comment on the Letter "Geometric Origin of the Tennis Racket Effect'' by P. Mardesic, et al, Phys. Rev. Lett. 125, 064301 (2020)}, arXiv:2302.04190.  


\end{thebibliography}
\end{document}